# Controlling plasma distributions as driving forces for ion migration during fs laser writing

Toney Teddy Fernandez[1], Jan Siegel[1,*], Jesus Hoyo[1], Belen Sotillo[2], Paloma Fernandez[2], Javier Solis[1]

[1] *Instituto de Óptica, C.S.I.C., Serrano 121, 28006 Madrid, Spain*

[2] *Depto. de Física de Materiales, Facultad de Físicas, Univ. Complutense, 28040-Madrid, Spain*

*Corresponding author: j.siegel@io.cfmac.csic.es

**Abstract**

The properties of structures written inside dielectrics with high repetition rate femtosecond lasers are known to depend strongly on the complex interplay of a large number of writing parameters. Recently, ion migration within the laser-excited volume has been identified as a powerful mechanism for changing the local element distribution and producing efficient optical waveguides. In this work it is shown that the transient plasma distribution induced during laser irradiation is a reliable monitor for predicting the final refractive index distribution of the waveguide caused by ion migration. By performing in-situ plasma emission microscopy during the writing process inside a La-phosphate glass it is found that the long axis of the plasma distribution determines the axis of ion migration, being responsible for the local refractive index increase. This observation is also valid when strong positive or negative spherical aberration is induced, greatly deforming the focal volume and inverting the index profile. Even subtle changes in the writing conditions, such as an inversion of the writing direction (quill writing effect), show up in the form of a modified plasma distribution, which manifests as a modified index distribution. Finally, it is shown that the superior control over the waveguide properties employing the slit shaping technique is caused by the more confined plasma distribution produced. The underlying reasons for this unexpected result are discussed in terms of non-linear propagation and heat accumulation.

## 1. Introduction

The concept of focusing ultrashort laser pulses inside dielectric materials to write optical waveguides for applications in Photonics has been introduced almost two decades ago [1,2,3]. With the advent of high repetition rate fs lasers, operating at several hundreds of kHz up to a few MHz, losses could be strongly decreased due to the combined action of ultrafast non-linear absorption and slower heat accumulation and diffusion processes [4,5]. Optimization of fs laser-written structures and devices is carried out basically by tuning the laser parameters. Hence the early strategy of this promising technique was largely based on empirically varying the laser-dependent parameters, which often resulted in uncontrolled structure dimensions and low refractive index change.

The reports somehow overlooked the existence of a material-dependent parameter related to the substrate composition, until it was demonstrated recently that controlled, fs-laser induced ion-migration can be used to produce highly efficient guiding structures with very large index contrast. This strategy enables overcoming the low refractive index change bottleneck by a proper design of the glass composition [6,7]. It was shown that laser irradiation led to local compositional changes via ion migration inside the glass, as reported first in static irradiation experiments [8,9]. Applied to waveguide writing, this mechanism has the ability of producing superior refractive index changes, in excess of $10^{-2}$. Using this technique very efficient active/passive waveguides have been demonstrated in in-house fabricated [7] and commercial phosphate [6] glasses with different glass modifiers. The refractive index changes induced were identified to be caused by the migration of heavy elements. Recently, Sakamura et al. investigated the possibility of controlling the shape of the elemental distributions inside the glass [10]. The authors used a spatial light modulator producing multiple foci and managed to change correspondingly shape of the elemental distribution. It has to be said, though, that these experiments were static and not aimed at increasing the local refractive index or fabricating optical waveguides.

In order to obtain a precise control over elemental redistribution during waveguide writing it seems therefore essential to control the intensity distribution of the focal volume of the laser beam inside the glass. Yet, the initial shape of the volume in which the laser energy is deposited (effective focal volume, EFV) is unknown and difficult to determine theoretically. Even for an apparently simple situation of focusing a circular Gaussian laser beam inside a material with a single lens poses complex problems to modeling attempts due to the combined presence of a) spherical aberration, b) non-linear propagation, c) non-linear absorption, d) thermal accumulation effects, and e) sample movement, which leads to a dynamically changing situation that depends also on the pulse energy laser repetition rate and sample translation speed. The entangled dynamic interaction of all these factors renders the problem even more complex. Additional problems are added when using slit-shaped laser beams to achieve a disk-like focal volume [11], in particular because of the astigmatism introduced, accompanied by truncation and diffraction of the beam at the slit [12].

Some works have attempted to address these problems using different modeling approaches. Hoyo et al. implemented a fast Fourier evolver, demonstrating its great potential in predicting the EFV in presence of spherical aberration, non-linear propagation and absorption in phosphate glasses, though not including heat accumulation effects as present with high repetition rate lasers [13,14]. Eaton et al. investigated experimentally and by means of calculations the effect of heat diffusion and accumulation for high repetition laser irradiation [4]. The model assumed an asymmetric laser-heated focal volume and spherical Gaussian heat dissipation. Despite being a static model (no sample movement) with an effective number of incident laser pulses, the model predicted waveguide diameters in good agreement with their experimental results. Ye et al. investigated, also in a static study, the effect of spherical aberration in combination with heat accumulation induced by high repetition rate lasers, yielding 2D temperature distributions that resembled reasonably well the shape of the spots written at different depth [15]. A thermal conduction model with a moving heat source was developed by Miyamoto et al. to simulate the dimensions of the region molten by the laser irradiation, starting from the experimentally measured fraction of absorbed energy, denominated non-linear absorptivity [16]. The most comprehensive modeling approach, to the best of our knowledge, was presented by Sun and co-workers, who complemented a non-linear propagation/absorption model by taking into account thermal diffusion [17]. In that way, the authors obtained good agreement with the experimental cross-sections of waveguides written in [16]. It has to be noted though that the structures written in that reference did not show waveguiding properties, in part due to the too long pulse duration used (10 ps).

Prompted by the high complexity for modeling the laser interaction mechanisms in presence of the above listed difficulties we approached the problem from an experimental perspective. Using plasma emission microscopy during waveguide writing we are able to obtain a snapshot of the 2D spatial distribution of the laser-induced plasma in true writing conditions. In other words, we image the actual EFV influenced by spherical aberration, non-linear propagation, non-linear absorption, thermal accumulation effects and sample movement as it is converted into a plasma. The so-obtained 2D emission images are compared with in-situ white light transmission images and DIC ex-situ microscopy images of the waveguides written. By spatially overlapping the different images, a precise correlation between deposited laser energy distribution and final index distribution is obtained. A similar approach, yet without precise spatial overlap of plasma and white light transmission images, was reported very recently in the study of microwelding of borosilicate glass [18].

The paper is structured as follows: After a brief description of the experimental configurations and conditions used, the results are presented, grouped in four subsections. The first two concentrate on results obtained upon waveguide writing with a circular beam and slit-shaped beam, respectively. In particular, the experimentally recorded 2D plasma distributions for different pulse energies are compared in terms of size, shape and position to the waveguide cross sections obtained with DIC microscopy in order to investigate the relation between the initially laser-excited volume and the ion migration direction and extension. In the third subsection, we introduce in a controlled way spherical aberration with positive or negative sign in order to shape the EFV and thus influence the ion migration direction. In the last subsection, the subtle influence of the writing direction on the plasma distribution and waveguide cross section is explored, contributing to the ongoing discussion about quill writing effects [19,20].

## 2. Materials and methods

We used phosphate glass samples made of Kigre-QxErSpa100 (composition ~ 65·$P_2O_5$-10·$La_2O_3$-10·$Al_2O_3$-10·$K_2O$ %mol.) doped with $Er_2O_3$ (2% wt.) and $Yb_2O_3$ (4% wt.). The laser used was a high repetition rate, femtosecond laser amplifier operating at a wavelength of 1030 nm and a repetition rate of 500 kHz. The pulse duration was 500 fs and the beam diameter before the focusing lens was 4.3 mm ($1/e^2$) with circular polarization. The measurement of the laser beam properties were performed using a GRENOUILLE device, which also provided a measurement of the pulse front tilt (PFT), yielding a value of 0.75 fs/mm. Optical waveguides were written inside the glass by focusing the beam, at normal incidence to the sample surface (along the z-axis), below the surface at depths ranging from 50 μm – 300 μm, using two different lenses. First, a 0.68 NA aspheric lens was used to fabricate waveguides with optimum performance in terms of minimal losses at 1640 nm and excellent mode matching to an SMF-28 fiber [6]. Second, a microscope objective lens (0.8 NA) with coverslip correction (0.17 mm) was employed to perform irradiations with different amounts of spherical aberration. Coverslip correction means that no spherical aberration is induced when focusing the laser beam at a depth of 170 μm, but leading to positive spherical aberration at higher depth and negative spherical aberration at lower depth. In some experiments the slit shaping technique [11,12] was employed by sending the laser beam through a narrow slit aligned parallel to the y-axis in order to widen the diameter along the x-axis of the focal volume. As variable processing parameters for the study we used the sample scan speed ($v$ = 40, 60 and 80 μm/s) moving along the y-axis either from -y to +y or +y to -y, the slit width (no slit, $s$ = 1.4 mm) and the laser energy (400 - 800 nJ). Further details on the writing set-up are given elsewhere [6,7].

An in-situ plasma microscopy setup was implemented, as sketched in Figure 1. It is composed of a long working distance

50x, 0.42 NA microscope objective lens and a 12 bit charge-coupled device (CCD) camera, installed along the y-axis in order to record side-view images of the plasma emission in the focal region. Microscopy images of the plasma emission were acquired while translating the sample along the x-axis, i.e. in true writing conditions. As a consequence of this configuration the plasma distribution appeared only during a very short time in focus. The depth of focus of the plasma microscope is DOF ≈ 4 μm, which means that for a sample translation speed of $v$ = 60 μm the plasma moves through the focus at $v_{plasma}$ = 60 μm/s / $n$ ≈ 40 μm/s, $n$ being the refractive index of the sample (n = 1.55 in the visible range). The plasma distribution is therefore in focus during a time $t$ = 100 ms. In order not to miss the image that is best focused, the camera recorded a video of the plasma distribution approaching the focal plane with a maximum frame rate of 10 images per second and a maximum exposure time of 80 ms. Since the intensity of the plasma emission changed by orders of magnitude over laser energy range explored, the exposure time had to be decreased at high energies (and in some cases neutral density filters used), in order to prevent saturation of the CCD. The related attenuation via exposure time or filter adjustment was subsequently corrected for in the recorded images.

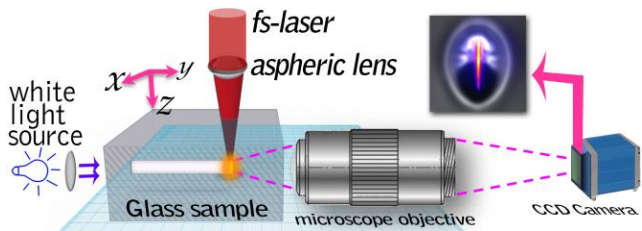

*Figure 1: Waveguide writing and in-situ plasma microscopy set-up*

The plasma microscopy setup could be switched easily into an in-situ white light transmission microscope for the waveguides written by switching on a collimated white light illumination source installed at the opposite side of the sample along the y-axis. By recording such an image of the waveguide cross section, directly after having recorded the corresponding plasma image, a perfect spatial overlap between both images was ensured. After merging both images, one in black and white and the other in false color scale, a precise localization of the plasma distribution within the waveguide region written is possible. For a final characterization, the sample was polished on both sides and images of the waveguide cross sections were recorded using transmission differential interference contrast (DIC) mode in a Nikon Eclipse optical microscope with a 100x, NA= 0.9 objective lens. An analysis of the local elemental composition of the waveguide cross sections was performed using energy-dispersive x-ray microanalysis with a Leica S440 SEM equipped with a Bruker AXS Quantax μ-analysis system, having a resolution of 125 eV.

## 3. Results and discussion

### 3.1. Plasma distribution and waveguide cross section for a circular writing beam

The first set of experiments consisted in recording plasma emission images upon focusing the circular fs laser beam operating at 500 kHz repetition rate with an aspheric lens (NA 0.68) at a depth of 100 μm below the surface of the sample while the latter was being translated at 60 μm/s along the y-axis. The upper row of Figure 2 shows images of x-z plane of the focal region with the laser beam being incident from the top (along the z-axis, c.f. Figure 1). The top left image shows the calculated linear intensity distribution using the model developed by Török and co-workers for computing spherical aberration effects [21]. In fact, the slight asymmetry observed along the propagation axis is caused by spherical aberration induced by the refractive index mismatch at the air/glass interface. This asymmetry can be seen more clearly in the corresponding intensity profiles in Figure 2.

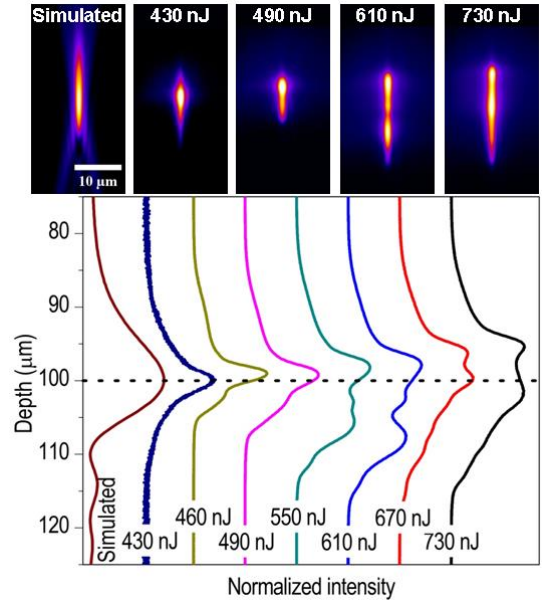

*Figure 2: The upper row shows side view images of the focal region of the laser beam (incident from the top, z-axis) focused with an aspheric lens (NA = 0.68) at a depth of 100 μm inside the glass sample. The top left image corresponds to the calculated intensity distribution, the other images ones to experimental plasma emission images recorded at different writing energies (see labels) while translating the sample at 60 μm/s along the y-axis. The plot below shows normalized longitudinal intensity profiles of calculated and experimental images along the z-axis.*

For comparison, experimentally recorded plasma emission images are shown in Figure 2. At the lowest writing energy (430 nJ), the plasma distribution has a quite similar size and shape to the calculated incident laser intensity distribution, considering that a complex non-linear relation exists between plasma intensity and laser excitation intensity distributions. This similarity indicates that there is only a small influence of non-linear and thermal accumulation effects at this low writing energy. Yet, as can be seen by comparison of the longitudinal profiles in Figure 2, the shoulders/oscillations after the peak visible in the calculated distribution are absent at 430 nJ. The possible reason is the existence of a threshold for plasma formation, which leads to a suppression of low intensity wings.

Upon increasing the writing energy, the shape of the plasma distribution changes strongly. At 490 nJ the distribution gets elongated in form of a sharp first peak and a longer tail. As the energy is further increased, the distributions get longer and feature multiple peaks with an overall complex profile. The beam filamentation is a clear sign for the presence of non-linear effects [22]. Moreover, the depth at which the peak of the plasma filament is located shifts towards the surface for increasing energies, consistent with a focal shift due to non-linear self-focusing [23]. It is worth noting that we found no evidence for periodic oscillations of the depth and shape of the plasma distribution as reported very recently by Miyamoto and co-workers [18]. They used a high-speed CCD camera with very short exposure times (5 μs), which allowed them to record oscillations with a period of a few kHz. While our CCD with an exposure time of 2 ms would not be able to resolve such oscillations, if present, we want to point out that for certain pulse energies and lower repetition rates we did observe periodic intensity oscillations with much longer period (typically 1 second). Yet we made sure to avoid this regime since the propagation losses of the resulting waveguides obtained were high due to the varying diameter of the waveguide cross section.

It is enlightening to compare the plasma images in terms of size and position to the cross sections of the waveguides written during the experiment. For this purpose we have recorded, using the plasma microscopy setup, a white light transmission image of the waveguides written, directly after recording the plasma image with the sample being in the same position and focal plane. This ensured a perfect spatial overlap between plasma and transmission images. Both images have been merged and are shown in Figure 3, with the plasma image being in false color scale and the transmission image in grayscale. It can be seen immediately that the plasma filament is located approximately in the center of the structure written, which has an inverted tear-drop shape. The thermal accumulation due to the high repetition rate of the laser leads to this considerable increase of the modified volume, starting from the narrow plasma streak. This is, to the best of our knowledge, the first time, that a precise experimental determination and overlap of initial plasma distribution and waveguide (cause and effect) has been performed.

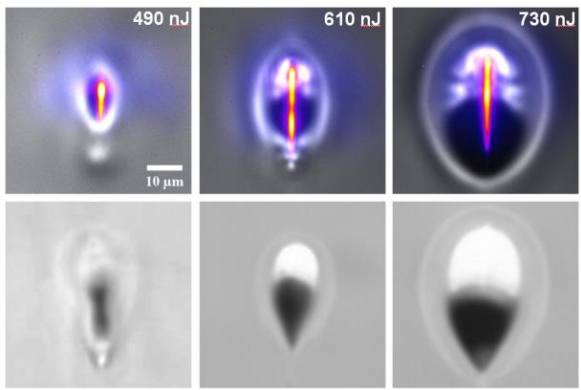

*Figure 3:* Side view images of the laser-modified region (laser incident from the top) at a depth of 100 μm inside the glass sample. The upper row shows merged in-situ images composed of trans-illumination images (grayscale) and plasma emission images (false color scale). The bottom row displays ex-situ recorded DIC micrographs of the same structures after polishing.

While the spatial extension of the waveguides can be appreciated in the in-situ trans-illumination images, their interpretation is compromised by the fact that the end facets of the sample were damaged due to surface ablation during the writing process. Therefore, the sample was polished on both sides and an ex-situ transmission differential interference contrast (DIC) micrograph of each structure was recorded. The results are included in Figure 3. At the lowest energy, mostly dark regions show up, indicative of a local decrease of the refractive index. The structures written at 610 nJ and 730 nJ feature also a bright region, indicative of a refractive index increase. These regions of high and low refractive index are surrounded by a low contrast grey region with an inverted tear-drop shape, marking the region of heat diffusion. Previous studies with EDX mapping on this sample under the same writing conditions have unambiguously demonstrated that a high refractive index increase ($> 10^{-2}$) in the bright region has been produced by ion migration, predominantly lanthanum (La) enrichment.[6] The accompanying potassium (K) depletion does, however, not contribute substantially to the index change [24].

Figure 4 shows a secondary electron image of the waveguide written at 730 nJ, evidencing a strong Z contrast, which is consistent with the enrichment of the guiding region with a heavy element and its depletion in the low refractive index zone. EDX measurements of the local composition of this waveguide along the z-direction are shown in the plot to the right in Figure 4, confirming local La enrichment to be responsible for the refractive index increase and Z contrast observed. It is worth noting that the plasma distribution is not confined to the region of La enrichment but extends towards the region of La depletion and K enrichment.

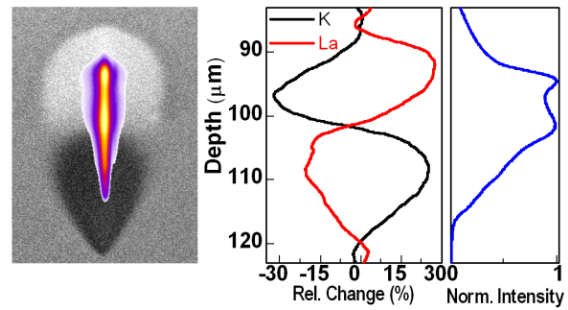

*Figure 4:* (left) Secondary electron image of the waveguide written at 730 nJ (c.f. Figure 3) superimposed by the plasma distribution measured. Results of EDX measurements are shown in the plot to the right compared with the intensity profile of the plasma distribution.

By comparing the secondary electron and DIC images of the polished waveguides to the merged composite images containing the plasma distributions we can confirm that the plasma distribution under the present conditions crosses the boundary between the two regions of low and high refractive index. This result indicates that the plasma distribution is the underlying driving force for ion migration. As for the longitudinal plasma profile (c.f. Figures 2 and 3), it can be observed that the intensity peak preceded by a sharp increase lies in the region that turns later into a high refractive index (La-enriched) compared to the low refractive index region (La-depleted) containing the decaying part of the plasma distribution.

In this context, a related study has been performed by Luo et al. in a silicate glass [25], investigating the influence of asymmetric longitudinal intensity profiles on ion migration. Based on their results they found that in the region of peak intensity an accumulation of certain network modifiers occurs, consistent with our finding of La accumulation. While their work relied on calculated intensity distribution and without the possibility of a spatial overlap to the experimental structures (non-guiding in their case) our results show that this hypothesis is correct also in our case under the present conditions. However, as will be shown in Section 3.3., this behavior is not a general rule and can change when processing conditions are modified.

## 3.2. Plasma distributions and waveguide cross sections for a slit-shaped beam

In a second set of experiments we investigated the way in which slit shaping of the laser beam influences the plasma distributions. The positive effect of this technique on the properties (waveguide cross section and propagation losses) of the waveguides produced in this glass has been shown previously [6,24]. What remained unclear was the observation that the waveguide diameter turned out to be much less sensitive to a change of the writing energy when using slit shaping and that waveguides with smaller diameter, yet high index contrast, could be written. It seems likely that the underlying reason is linked to the laser-induced plasma distribution.

Figure 5 shows the results equivalent to Figure 2 for a slit shaped beam. As for the calculated intensity distribution, the main difference is found in a larger transversal diameter (y-axis) in the focal region as expected for a smaller input beam diameter in this direction due to the slit confinement (similarity theorem of the Fourier transform). Only little influence is found in the longitudinal direction (c.f. corresponding profile in Figure 5), slightly reducing the amplitude of the shoulders due to the reduced effective numerical aperture in the x-axis.

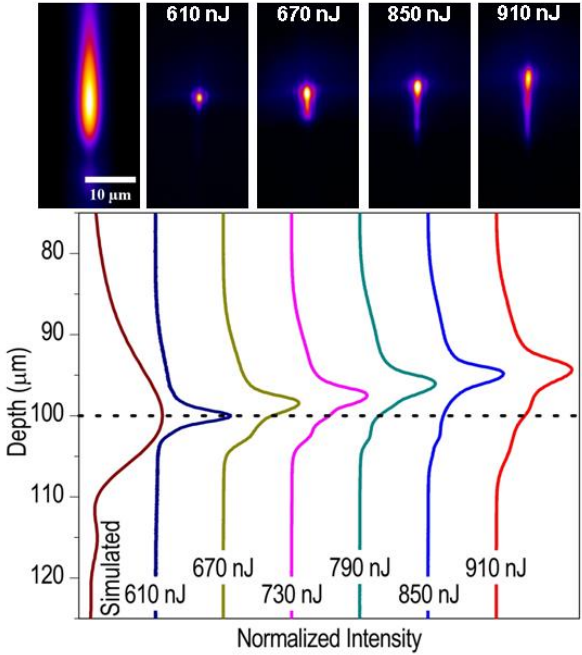

*Figure 5:* The upper row shows side view images of the focal region of a slit-shaped laser beam (incident from the top) focused with an aspheric lens (NA = 0.68) at a depth of 100 µm inside the glass sample. The top left image corresponds to the calculated intensity distribution, the other ones to experimental plasma emission images recorded at different writing energies (see labels) while translating the sample at 60 µm/s along the y-axis. The plot below shows normalized longitudinal intensity profiles of calculated and experimental images along the laser propagation direction (z-axis).

Based on this calculation one would expect plasma distributions similar to those without slit shaping, with slightly larger transversal dimension. Yet, as shown in Figure 5 the distributions are notably different to those shown in Figure 2. As can be clearly appreciated by a comparison of the corresponding longitudinal profiles, the slit-shaped profiles recorded at different energies are considerably shorter and, importantly, very similar to each other, despite the wider energy range explored compared to the study without slit.

We have also analyzed the transversal plasma profiles. The experimental ones for a slit-shaped beam turn out to be significantly narrower ($d_{x,slit}$ = 1.1 – 1.6 µm, FWHM diameter) than predicted from the calculated linear intensity profile ($d_{x,slit\_calc}$ = 4.6 µm). In fact their diameters are comparable to those obtained without slit shaping ($d_{x,no-slit}$ = 1.2 – 1.8 µm). This is a clear indication for non-linear and/or thermal effects conditioning the effective focal volume. Possible mechanisms involved in this respect are self-focusing [23] and/or thermal lensing [26,27]. The presence of these mechanisms can also be appreciated in form of a gradual shift of the focal depth in the profiles in Figure 5, initially at 100 µm at low energy, to significantly smaller values (92 µm) at high energy.

Based on the measurements of longitudinal and transversal cross sections of the plasma distributions we have determined their aspect ratio as AR = $d_z/d_x$ and area as A = $\pi \cdot d_z \cdot d_x/4$. The results are shown in Figure 6. It can be immediately seen that both aspect ratio and area increase strongly above a threshold energy (500 nJ) in the case of a circular beam. An excessive aspect ratio as obtained here, with values up $AR_{no\ slit}$ = 10, is obviously adverse for generating waveguides with circular cross sections. Similarly, plasma distributions with a too large areas (up to $A_{no\ slit}$ = 17 µm$^2$) are also adverse for waveguide writing as these lead to waveguides with too large diameters for single mode propagation. In contrast, for a slit-shaped beam both the aspect ratio and area value remain low and approximately constant ($AR_{slit}$ ≈ 3 and $A_{slit}$ ≈ 5 µm$^2$).

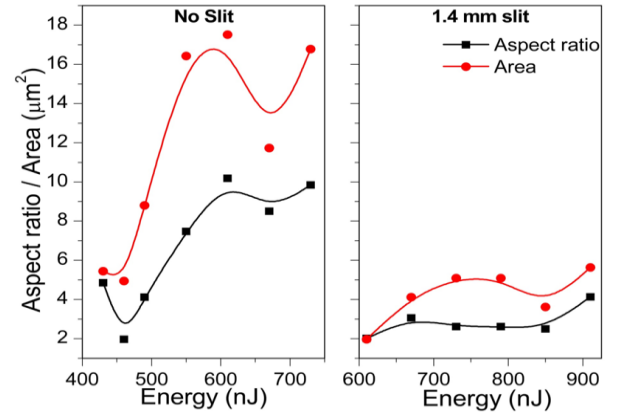

*Figure 6:* Evolution of aspect ratio and area of the experimentally measured plasma distributions shown in Figures 2 and 5 with pulse energy for a circular input beam ("no slit", left) and a slit-shaped beam ("slit", right). Please note that both plots have the same vertical scale.

While the ideal aspect ratio of AR = 1 is not reached under our conditions, moderate aspect ratios in the plasma distributions can be compensated for by thermal heat flow to lead to waveguides with a further reduced aspect ratio as shown in [24]. These results explain findings reported in previous works using slit shaping [6,24], showing that the superior control over the waveguide properties obtained with slit shaping is caused by the spatially confined plasma distributions. In particular, the evolution of the diameter of the guiding region with pulse energy reported in [24] shows a very similar behavior as the corresponding curves for the plasma distribution shown in Figure 6, both for a circular and a slit shaped beam.

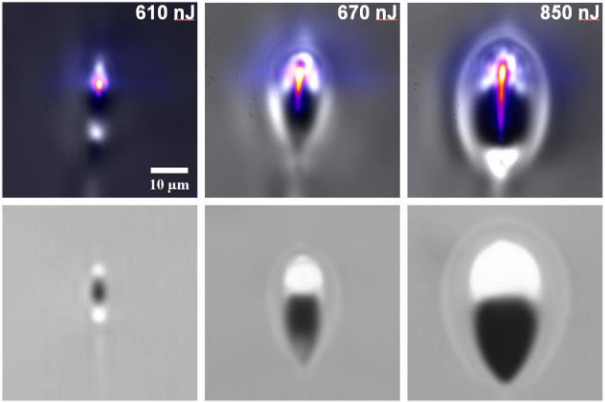

*Figure 7: Side view images of the laser-modified region (laser incident from the top) at a depth of 100 µm inside the glass sample. The upper row shows merged in-situ images composed of trans-illumination images (grayscale) and plasma emission images (false color scale). The bottom row displays ex-situ recorded DIC micrographs of the same structures after polishing.*

The superposition of plasma images and trans-illumination images is shown in Figure 7. Essentially as for a circular beam, the plasma distributions are located at the surface-near region of the structure written, for all writing energies used. Also here, a comparison of the merged images to the DIC images taken of the polished waveguides reveals that in all cases the peak of the plasma distribution lies in the region of increased refractive index (bright region) and its tail extends in to the region of depressed index (dark region).

## 3.3. Controlling spherical aberration (SA) for shaping the focal volume

The results shown in the previous sections demonstrated that for a longitudinally asymmetric plasma distribution the high index region is typically located near the peak of the plasma distribution and the depressed index region near its long tail. In the present section we follow a similar experimental approach as the one used by Luo et al. [25], introducing spherical aberration (SA) in order invert the longitudinal intensity distribution of the focal volume and thus the index distribution. While the above-cited work used an oil immersion objective lens and index matching liquids with different refractive indices, we preferred to choose a different approach in order to avoid the use of oil immersion lenses, which would impose important restrictions for 3-D optical waveguide writing. Our strategy, detailed in the experimental section, is based on using a dry objective lens with relatively high NA, which is corrected in terms of SA for the presence of a 170 µm thick coverslip. A simple way to introduce negative SA and thus to invert the longitudinal intensity profile is to write at shallower depth.

Figures 8 (a,b) show the calculated intensity distribution and profile for the case of a circular beam (no slit) focused with a coverslip-corrected NA = 0.85 lens at a depth d = 50 µm inside the sample. The inversion of the intensity profile compared to the profile shown in Figure 2 is evident. The higher number of oscillations compared to Figure 2 is due to the higher NA used, on which SA depends strongly.

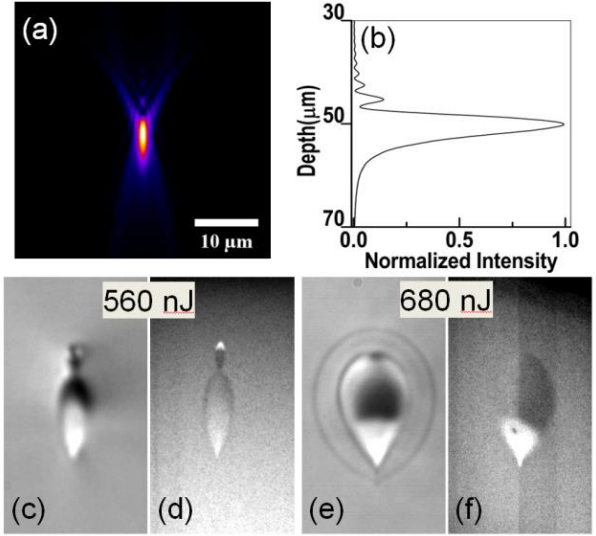

*Figure 8: (a) Side view image and (b) corresponding profile of the calculated intensity distribution of the focal region of a circular laser beam (incident from the top) with a coverslip-corrected dry objective lens (NA = 0.85) at a depth of 50 µm inside the glass sample. (c,e) DIC micrographs and (d,f) secondary electron images of optical waveguides written under these conditions, after polishing. All images have the same scale, given by the scale bar in (a).*

Using this lens and writing depth we have written optical waveguides at the same writing speed and laser repetitions rate as for Figure 2. Figures 8(c,e) show DIC images of two optical waveguides written with different energies. While both images feature a quite small size of the guiding (white) region, the important feature to observe is that the longitudinal refractive index profile has been inverted compared to Figures 3 and 5. This is confirmed by SEM measurements displayed in Figures 8(d,f), showing secondary electron images of the waveguide written. In both images the Z contrast distribution is inverted with respect to Figure 4. These results seem to indicate that an inversion of the intensity profile leads to an inversion of the ion migration direction. Yet, this conclusion would be based on a very simplified calculation of the intensity profile and it is therefore necessary to measure the EFV using in-situ plasma microscopy.

Figure 9 provides an overview of the plasma emission, trans-illumination and DIC microscopy study performed introducing controlled spherical aberration in order to modify the focal volume, namely negative SA (d = 50 µm depth), positive SA (d = 300 µm depth) and negligible SA (d = 100 µm). It is worth noting that we have determined experimentally the depth at which negligible SA is introduced rather than relying on the value specified on the objective lens (170 µm). The reason is that these specifications are provided for use with a different glass with different refractive index and under linear propagation conditions. In turn, our strategy for experimental determination of SA = 0 was based on measuring the threshold energy for plasma emission at different depth, which should be minimum for SA = 0 and which was in our case identified as d = 100 μm.

At d = 50 μm (negative SA), plasma microscopy reveals an unexpected result, as can be seen in the left column of Figure 9. While the main plasma peak is indeed preceded by a single smaller peak towards the surface, as expected for negative spherical aberration, both are accompanied by a tail extending inwards into the sample. Also, these two main peaks are not located as before in the guiding region but in the index-depressed region. These two observations lead to the conclusion that the ion migration direction can indeed be inverted by modifying the EFV, but that the process is not as simple as assumed.

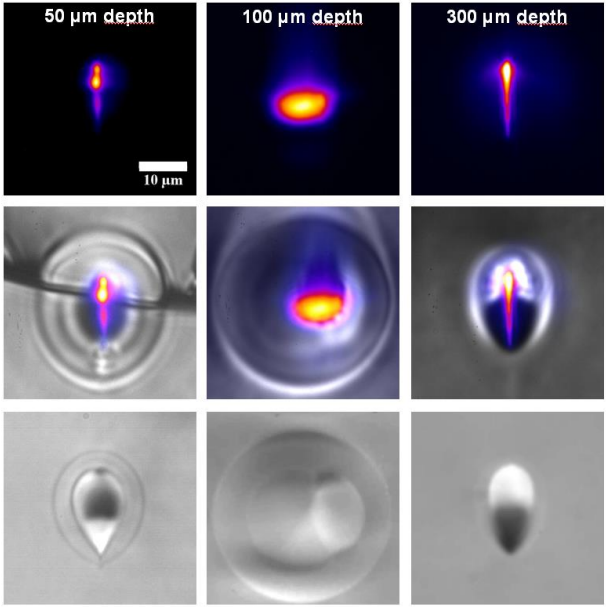

*Figure 9:* Side view images of the laser-modified region (laser incident from the top) at different depths inside the glass sample. Left column d= 50 μm, 680 nJ, middle column d= 100 μm, 720 nJ, right column d= 300 μm, 800 nJ, all written at 60 μm/s. The upper row shows plasma emission images, the middle row merged in-situ images composed of trans-illumination images (grayscale) and plasma emission images (false color scale) and the lower row displays ex-situ recorded DIC micrographs of the same structures after polishing.

This conclusion is found valid also for the results obtained at d = 100 μm (SA = 0), displayed in the middle column of Figure 9. At this depth, the absence of spherical aberration leads to an intensity profile that is longitudinally symmetric and compressed. However, it is transversally elongated, most likely due to non-linear propagation and absorption, yielding a rotated plasma profile with an inverted longitudinal/transversal aspect ratio compared to the previous results. The direct consequence of this plasma distribution rotated by 90° is a rotated index distribution, as can be seen in the corresponding DIC image. Also, it can be appreciated in the merged image that the position of the plasma distribution lies, again, not in the high index region. Instead, in this case it is approximately centered at the boundary between the regions of increased and depressed index.

The results obtained at d = 300 μm (positive SA) are shown in the right column of Figure 9. They are very similar to the results obtained in the previous sections using a lower NA lens with the presence of positive SA. This consistency confirms that SA can indeed be used to control the plasma distributions and, in that way, the ion migration direction and final index profiles of the waveguides. In this respect it is worth noting that in terms of ion migration and waveguiding properties, the best structures are achieved using positive SA.

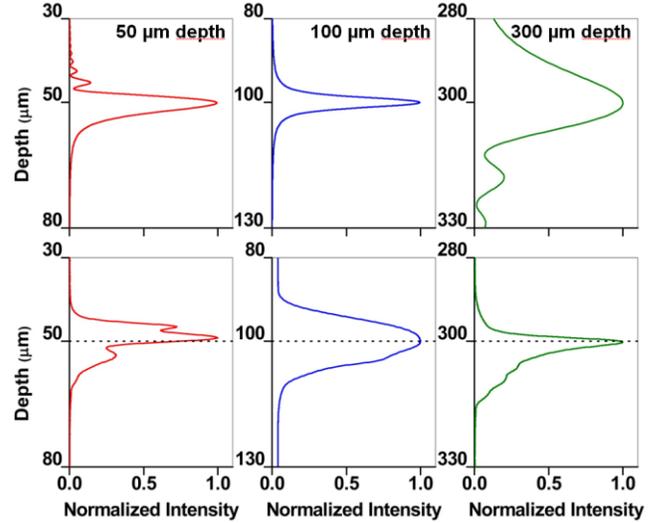

*Figure 10:* Longitudinal intensity profiles for the results obtained at different depths shown in Figure 9, extracted from (upper row) calculations considering spherical aberration and (lower row) experimentally measured plasma intensity profiles.

Figure 10 shows a comparison of the calculated longitudinal intensity profiles considering spherical aberration with the experimentally measured plasma intensity profiles. Although far from being an exact match, due to the over-simplicity of the model neglecting non-linear propagation and thermal accumulation, the expected symmetry for 100 μm depth and asymmetry for 50 μm and 300 μm is reflected in the experimental results.

### 3.4. Influence of the writing direction: Quill writing

As a last point we have investigated the influence of the writing direction on the plasma and final index distribution, i.e. moving the sample left-to-right (+y) or right-to-left (-y). The influence of this parameter on the symmetry of the written structures pulses has been reported first by Kazansky and coworkers [19,20], who coined the term "quill writing". They provided evidence that the directional dependence is caused by a pulse front tilt (PFT) of the laser pulses, present in many lasers with diffractive or dispersive components and often caused by a non-perfect alignment of the stretcher or compressor [28]. This discovery triggered strong interest in the community since it provided a simple, previously-overlooked experimental parameter for writing structures with different properties.

More recently, Poumellec et al. proposed a different underlying mechanism for the quill writing effect [29]. Their interpretation is based on the ponderomotive force induced by the light field and acting on the plasma distribution, thus generating transversal asymmetry. In this sense our experimental approach of performing plasma microscopy during writing is ideal in order to shed light on the underlying mechanisms. Moreover, most studies have investigated the influence of the writing direction on laser-written birefringent regions, nanogratings, bubbles, etc. [19] while optical waveguides have been neglected. Our concern here is to study the influence of the quill writing effect on the waveguide properties.

We have performed the same experiment as shown in Figure 9 for a depth of d = 100 μm (SA = 0) upon writing one waveguide left-to-right (forth, standard direction used in the rest of the paper) and another right-to-left (back). The result is shown in Figure 11. Consistent with the result shown in Figure 9, at this depth plasma distribution is elongated in the x-direction, leading to a refractive index gradient in this direction. However, upon inversion of the writing direction and yet keeping all other parameters constant, the plasma distribution changes notably. It becomes more circular, which expresses in the corresponding DIC image in a reduced index gradient, which almost vanishes.

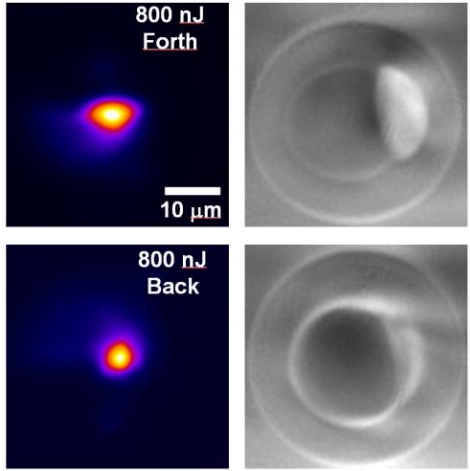

*Figure 11: Side view images of laser-modified regions (laser incident from the top) at d = 100 μm inside the glass sample. Top row: Writing direction left-to-right (+y, "forth"). Bottom row: Writing direction right-to-left (-y, "back"). The first column shows plasma emission images, the second column DIC micrographs of the same structures after polishing.*

The conclusions that can be drawn from these results are manifold. First, the long axis of the plasma distribution marks the direction of ion migration. This can be considered as a general conclusion for all results shown in this paper, without exception. Second, the strongly reduced index contrast obtained for an almost circular plasma distribution indicates that an AR = 1 is not ideal and that higher aspect ratio values (e.g. AR = 3, c.f. Figure 6) yield to a larger refractive index increase. This conclusion is consistent with the underlying concept of ion migration, which should yield to a stronger element separation for linear migration (as caused by an elongated plasma distribution with an AR > 1) than for radial migration (caused by a spherical plasma distribution with AR = 1). And third, our experimental results demonstrate that the transversal shape of the plasma distribution is indeed changed by inverting the writing direction, as proposed by Poumellec [29]. Yet, this does not necessarily imply that this effect is caused by the ponderomotive force acting on the plasma. As for the possible influence of the PFT on the non-reciprocal behavior we have performed PFT measurements as explained in the experimental section. The value extracted from the data is PFT = 0.75 fs/mm. This value can be considered as high, when comparing it to the values reported by the Kazansky group (PFT = 0.04 – 0.08 fs/mm) in an experiment that showed clear directional dependence [19]. Therefore, the considerable PFT in our experiment could also be the responsible (or a contributing) factor for the observed quill effect.

## 4. Conclusions

We have investigated the role of the laser-induced plasma distribution on the ion migration effects responsible for waveguide formation inside a La-phosphate glass upon high repetition rate fs-laser irradiation. Performing in-situ plasma emission microscopy in true writing conditions, our results show that the plasma distribution is a reliable monitor for predicting the refractive index and thus the element distribution of the final waveguide produced via ion migration. In particular the results explain findings reported in previous works using slit shaping, showing that the superior control over the waveguide properties obtained with slit shaping is caused by the spatially confined plasma distributions. Moreover, the long axis of the plasma distribution is found to determine the axis of migration. Rotation or inversion of the plasma distribution is feasible by adjusting experimental parameters, ultimately enabling direct control over the ion-migration and index distribution. The key experimental parameters identified for enabling this control are laser energy, spherical aberration and writing direction. We consider these findings to be of great importance not only for waveguide writing but other applications involving fs-laser induced ion-migration.

**Funding sources and acknowledgments.**

This work was partially supported by the Spanish Ministry Economy and Competitiveness (MINECO, TEC2011-22422, MAT2012-31959), J.H. and T.T.F. acknowledge funding from the JAE CSIC Program (pre- and post-doctoral fellowships, respectively, co-funded by the European Social Fund). B. Sotillo acknowledges her funding in the frame of CSD2009-00013 (MINECO).

**References**

1. K. M. Davis, K. Miura, N. Sugimoto, and K. Hirao, "Writing waveguides in glass with a femtosecond laser" Opt. Lett. 21, 1729 (1996).
2. M. Ams, G. D. Marshall, P. Dekker, J. A. Piper and M. J. Withford, "Ultrafast laser written active devices" Laser & Photon. Rev. 3, 535 (2009)
3. K C Vishnubhatla, S Venugopal Rao, R Sai Santosh Kumar, R Osellame, S N B Bhaktha, S Turrell, A Chiappini, A Chiasera, M Ferrari, M Mattarelli, M Montagna, R Ramponi, G C Righini7, and D Narayana Rao, "Femtosecond laser direct writing of gratings and waveguides in high quantum efficiency erbium-doped Baccarat glass", J. Phys. D: Appl. Phys. 42 205106 (2009).
4. S.M. Eaton, H. Zhang, M.L. Ng, J. Li, W.-J. Chen, S. Ho, and P.R. Herman, "Transition from thermal diffusion to heat accumulation in high repetition rate femtosecond laser writing of buried optical waveguides" Opt. Exp. 16, 9443 (2008).
5. R. Osellame et al. (eds.), Femtosecond Laser Micromachining: Photonic and Microfluidic Devices in Transparent Materials, Topics in Applied Physics 123, Springer-Verlag Berlin Heidelberg (2012).

6. T. Toney Fernandez, P. Haro-González, B. Sotillo, M. Hernandez, D. Jaque, P. Fernandez, C. Domingo, J. Siegel, and J. Solis, "Ion migration assisted inscription of high refractive index contrast waveguides by femtosecond laser pulses in phosphate glass" Opt. Lett. 38, 5248 (2013).
7. J. Hoyo, V. Berdejo, T. Toney Fernandez, a Ferrer, a Ruiz, J. a Valles, M. a Rebolledo, I. Ortega-Feliu, and J. Solis, "Femtosecond laser written 16.5 mm long glass-waveguide amplifier and laser with 5.2 dB cm$^{-1}$ internal gain at 1534 nm" Laser Phys. Lett. 10, 105802 (2013).
8. Y. Yonesaki, K. Miura, R. Araki, K. Fujita, and K. Hirao, "Space-selective precipitation of non-linear optical crystals inside silicate glasses using near-infrared femtosecond laser" J. Non. Cryst. Solids 351, 885 (2005).
9. Y. Liu, M. Shimizu, B. Zhu, Y. Dai, B. Qian, J. Qiu, Y. Shimotsuma, K. Miura, K. Hirao, "Micromodification of element distribution in glass using femtosecond laser irradiation" Opt. Lett. 34, 136 (2009).
10. M. Sakakura, T. Kurita, M. Shimizu, K. Yoshimura, Y. Shimotsuma, N. Fukuda, K. Hirao, K. Miura, "Shape control of elemental distributions inside a glass by simultaneous femtosecond laser irradiation at multiple spots" Opt. Lett. 38, 4939 (2013).
11. Y. Cheng, K. Sugioka, K. Midorikawa, M. Masuda, K. Toyoda, M. Kawachi, and K. Shihoyama, "Control of the cross-sectional shape of a hollow microchannel embedded in photostructurable glass by use of a femtosecond laser" Opt. Lett. 28, 55 (2003).
12. A. Ruiz De la Cruz, A. Ferrer, J. del Hoyo, J. Siegel and J. Solis, "Modeling of astigmatic-elliptical beam shaping during fs-laser waveguide writing including beam truncation and diffraction effects" Applied Physics A 104, 687 (2011).
13. J. Hoyo, M. Galvan-Sosa, A. Ruiz de la Cruz, E. J. Grace, A. Ferrer, J. Siegel, J. Solis, "Modeling of single pulse 3-D energy deposition profiles inside dielectrics upon fs laser irradiation with complex beam wavefronts" Proc. SPIE 9131, 91310G 1 (2014).
14. J. del Hoyo, A. Ruiz de la Cruz, E. Grace, A. Ferrer, J. Siegel, A. Pasquazi, G. Assanto, and J. Solis, "Rapid assessment of nonlinear optical propagation effects in dielectrics", submitted to Sci. Rep. (2014).
15. D. Ye, Y. Guang-Jun, W. Guo-Rui, M. Hong-Liang, Y. Xiao-Na, and M. Guo-Hong, Chin. "The effect of spherical aberration on temperature distribution inside glass by irradiation of a high repetition rate femtosecond pulse laser" Chin. Phys. B 21, 025201 (2012).
16. I. Miyamoto, K. Cvecek, and M. Schmidt, "Evaluation of nonlinear absorptivity in internal modification of bulk glass by ultrashort laser pulses" Opt. Exp. 19, 10714 (2011).
17. M. Sun, U. Eppelt, W. Schulz, and J. Zhu, "Role of thermal ionization in internal modification of bulk borosilicate glass with picosecond laser pulses at high repetition rates" Opt. Mat. Exp. 3, 1716 (2013).
18. I. Miyamoto, Y. Okamoto, R. Tanabe, Y. Ito, "Characterization of Plasma in Microwelding of Glass Using Ultrashort Laser Pulse at High Pulse Repetition Rates" Phys. Proc. 56, 973 (2014).
19. W. Yang, P.G. Kazansky, Y. Shimotsuma, M. Sakakura, K. Miura, K. Hirao, "Ultrashort-pulse laser calligraphy" Appl. Phys. Lett. 93, 171109 (2008).
20. W. Yang, P.G. Kazansky, Y.P. Svirko, "Non-reciprocal ultrafast laser writing" Nat. Phot. 2, 99 (2008).
21. P. Török, P. Varga, Z. Laczik, and G. R. Booker., "Electromagnetic diffraction of light focused through a planar interface between materials of mismatched refractive indices: an integral representation" J. Opt. Soc. Am. A 12, 325 (1995).
22. L. Sudrie, A. Couairon, M. Franco, B. Lamouroux, B. Prade, S. Tzortzakis, and A. Mysyrowicz, "Femtosecond Laser-Induced Damage and Filamentary Propagation in Fused Silica" Phys. Rev. Lett. 89, 186601 (2002).
23. A. Brodeur, and S. L. Chin, "Ultrafast white-light continuum generation and self-focusing in transparent condensed media" J. Opt. Soc. Am. B 16, 637 (1999).
24. J. del Hoyo, R. Martinez-Vazquez, B. Sotillo, T.T. Fernandez, J. Siegel, P. Fernandez, R. Osellame, and J. Solis, "Control of waveguide properties by tuning femtosecond laser induced compositional changes" Appl. Phys. Lett. 105, 131101 (2014).
25. F. Luo, J. Song, X. Hu, H. Sun, G. Lin, H. Pan, Y. Cheng, L. Liu, J. Qiu, Q. Zhao, and Z. Xu, "Femtosecond laser-induced inverted microstructures inside glasses by tuning refractive index of objective's immersion liquid" Opt. Lett. 36, 2125 (2011).
26. S. J. Sheldon, L. V. Knight, and J. M. Thorne, "Laser-induced thermal lens effect: a new theoretical model" Appl. Opt. 21, 1663-1669 (1982).
27. M. Sakakura, M. Terazima, Y. Shimotsuma, K. Miura, and K. Hirao, "Heating and rapid cooling of bulk glass after photoexcitation by a focused femtosecond laser pulse" Opt. Express 15, 16800-16807 (2007).
28. S Akturk, M Kimmel, P O'Shea, R Trebino, "Measuring pulse-front tilt in ultrashort pulses using GRENOUILLE" Opt. Exp. 11, 491 (2003).
29. B. Poumellec, M. Lancry, R. Desmarchelier, E. Hervé, F. Brisset, and J.C. Poulin, "Asymmetric orientational writing in glass with femtosecond laser irradiation" Opt. Mat. Exp. 3, 1586 (2013).